\documentclass[aps,prl,twocolumn,floatfix,amsmath,superscriptaddress,amssymb,bibnotes]{revtex4-1}

\usepackage{graphicx}
\usepackage{dcolumn}
\usepackage{bm}
\usepackage{epstopdf}
\usepackage[]{SIunits}
\usepackage{units}
\usepackage{braket}
\usepackage{pbox}
\usepackage[colorlinks, linkcolor = blue, citecolor = blue, filecolor = black, urlcolor = blue]{hyperref}
\usepackage{xspace} % add trailing space after commands like \gpe

\def\beqa{\begin{eqnarray}}
\def\eeqa{\end{eqnarray}}
\def\nD{\mathcal{N}}

\def\psiSS{\psi_{dd}}

\def\br{{\bf r}}

\def\s{\ensuremath{\sigma}\xspace}
\newcommand{\figref}[1]{Fig.~\ref{#1}}

\def\O{\ensuremath{\Omega}\xspace}
\def\vOneD{\ensuremath{V_{1D}\xspace}}

\def\GamP{\ensuremath{\gamma_{e}\xspace}}
\def\GamDph{\ensuremath{\Gamma_{}\xspace}}
\def\relative{\ensuremath{z\xspace}}

\begin{document}

\title{Dipolar dephasing of Rydberg $D$-state polaritons}
\author{C. Tresp}
\email[]{c.tresp@physik.uni-stuttgart.de}
\affiliation{5. Phys. Institut and Center for Integrated Quantum Science and Technology}
\author{P. Bienias}
\affiliation{Institute for Theoretical Physics III and Center for Integrated Quantum Science and Technology, Universit\"{a}t Stuttgart, Pfaffenwaldring 57, 70569 Stuttgart, Germany}
\author{S. Weber}
\affiliation{Institute for Theoretical Physics III and Center for Integrated Quantum Science and Technology, Universit\"{a}t Stuttgart, Pfaffenwaldring 57, 70569 Stuttgart, Germany}
\author{H. Gorniaczyk}
\affiliation{5. Phys. Institut and Center for Integrated Quantum Science and Technology}
\author{I. Mirgorodskiy}
\affiliation{5. Phys. Institut and Center for Integrated Quantum Science and Technology}
\author{H. P. B\"{u}chler}
\affiliation{Institute for Theoretical Physics III and Center for Integrated Quantum Science and Technology, Universit\"{a}t Stuttgart, Pfaffenwaldring 57, 70569 Stuttgart, Germany}
\author{S. Hofferberth}
\email[]{s.hofferberth@physik.uni-stuttgart.de}
\affiliation{5. Phys. Institut and Center for Integrated Quantum Science and Technology}

% \date{\today}

%=================%
%    Abstract     %
%=================%

\begin{abstract}
We experimentally study the effects of the anisotropic Rydberg-interaction on $D$-state Rydberg polaritons slowly propagating through a cold atomic sample. We observe the interaction-induced dephasing of Rydberg polaritons at very low photon input rates into the medium. We develop a model combining the propagation of the two-photon wavefunction through our system with nonperturbative calculations of the anisotropic Rydberg-interaction to show that the observed effect can be attributed to pairwise interaction of individual Rydberg polaritons at distances larger than the Rydberg blockade.
\end{abstract}

%\pacs{Valid PACS appear here}
\maketitle

%=================%
%  Introduction   %
%=================%
Long-range and spatially anisotropic dipole-dipole (DD) interactions enable new approaches for preparing and exploring strongly correlated quantum systems \cite{Pfau2009b}. Magnetic DD interaction couples individual nuclear spins to nitrogen-vacancy centers in diamond \cite{Lukin2007b,Wrachtrup2008} and is observed in dipolar gases of ultracold atoms \cite{Pfau2007,Ferlaino2014}. Electric DD interaction determines the long-range interaction between polar molecules \cite{Ye2013} or Rydberg atoms \cite{Pillet2010,Saffman2010} and may allow to investigate phenomena such as quantum magnetism \cite{Rey2011,Zoller2015b,Pohl2015} and topological phases \cite{Lukin2012d} in these systems. Recently the angular dependence of the DD interaction between single Rydberg atoms has been fully mapped \cite{Browaeys2015b}. Here we study, for the first time, the anisotropic DD interaction between slowly travelling polaritons coupled to a Rydberg $D$-state via electromagnetically-induced transparency (EIT) \cite{Fleischhauer2000}.

Rydberg-EIT has emerged as a powerful approach to realizing few-photon optical nonlinearities \cite{Kurizki2005,Adams2010,Lukin2011,Vuletic2012}. This novel technique enables a variety of optical quantum information applications such as highly efficient single-photon generation \cite{Kuzmich2012b}, entanglement generation between light and atomic excitations \cite{Kuzmich2013}, single-photon all-optical switches \cite{Duerr2014} and transistors \cite{Hofferberth2014,Rempe2014b}, and interaction-induced photon phase shifts \cite{Grangier2012}. Additionally, it allows us to probe novel phenomena such as attractive interaction between single photons \cite{Vuletic2013b}, crystallization of photons \cite{Fleischhauer2013}, or photonic scattering resonances \cite{Buechler2014}. The electric DD interaction between pairs of Rydberg atoms has been studied extensively in the perturbative van-der-Waals regime \cite{Cote2005,Walker2008,Zoller2015}. Rydberg-EIT experiments have so far mostly employed Rydberg $S$-states, where the interaction has only very weak angular dependence. Recent experiments simultaneously prepare Rydberg atoms in $S$- and $P$-states \cite{Adams2013} or two different $S$-states \cite{Weidemueller2013b,Hofferberth2014,Rempe2014b}. The interaction between these energetically well-separated states enables novel entanglement schemes for atomic systems \cite{Fortagh2014} and increased flexibility in the manipulation of weak light fields \cite{Lesanovsky2014,Wu2015}.

In this work, we employ Rydberg EIT to investigate the anisotropic interaction between individual Rydberg polaritons. We observe that $D$-state Rydberg polaritons are decoupled from the EIT control field due to their pairwise interaction. We show that this effect is significant for incident probe photon rates corresponding to a mean spacing between polaritons larger than the Rydberg blockade distance. We extract the scaling of the decoupling rate with EIT control Rabi frequency and principal quantum number of the Rydberg $D$-state. Finally, we present a model which treats the complicated pair interaction as a dephasing process. Combining dephasing rates extracted from nonperturbative pair potential calculations with numerical propagation of the two-polariton wavefunction we model our experimental observations.
%
%=================%
%  Figure 1       %
%=================%
\begin{figure}
\begin{center}
\includegraphics{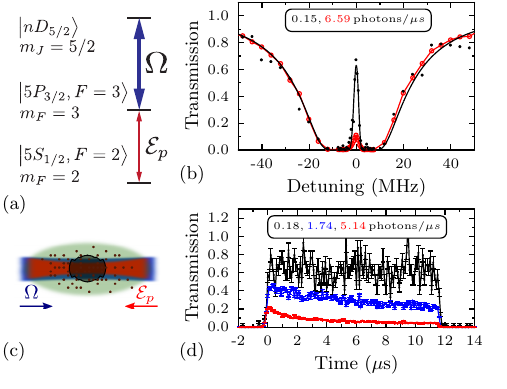}
\caption{\label{figure1} ~(a) EIT level scheme with week probe field  $\mathcal{E}_p$ and strong control field  $\Omega$. (b) Frequency scan of the weak probe field $\mathcal{E}_p$ showing the EIT transmission window averaged over the full pulse time. The transmission shows a strong nonlinearity with incident photon rate.  (c) Geometric scheme of our setup consisting of EIT lasers at $\unit[780] \nano \metre$ ($w_0 \approx \unit[6] \mu \metre$) and $\unit[480] \nano \metre$ ($w_0 \approx \unit[12] \mu \metre$) focused onto a thermal cloud ($\unit[30] \mu \kelvin$) of $^{87}Rb$ atoms with extensions of $\sigma_z=\unit[80] \mu \metre$ and $\sigma_r=\unit[25]\mu \metre$. The grey shadow shows the anisotropic blockade region caused by a $|100D_{5/2}\rangle$ Rydberg excitation. (d) Time dependent transmission on two-photon resonance for different probe photon input. }
\end{center}
\end{figure}
%
%=================%
%Experimental Part%
%=================%
Our EIT scheme consists of a few photon probe field $\mathcal{E}_p$ and a strong coupling field $\Omega$ resonantly coupling the levels illustrated in Fig.\,\ref{figure1}(a). Resulting transmission spectra of $\mathcal{E}_p$ for a frequency scan of the probe laser over EIT resonance employing the $|80D_{5/2}\rangle$ Rydberg state for two different photon rates $R_{in}$ are shown in Fig.\,\ref{figure1}(b). For $R_{in} = 0.15\, \text{photons}/ \mu \second$ we observe high transmission and a narrow linewidth, from which we extract a decoherence rate $\gamma_{gr}/2\pi = \unit[200] \kilo \hertz$, originating from the movement of the atoms due to the finite temperature and stray electric and magnetic fields over the cloud. For higher probe input rates the EIT transmission decreases, due to the self-blockade of propagating polaritons as observed in experiments using Rydberg $S$-states \cite{Adams2010, Vuletic2012, Kuzmich2012b}.
In contrast, here we observe a qualitatively new effect present only when we apply the EIT scheme to Rydberg $D$-states, namely a time dependence of the transmission on EIT resonance (Fig.\,\ref{figure1}(d)). For very low photon rates on the order of $R_{in}=0.15 \, \text{photons}/\unit \mu \second$, where we rarely have two photons inside the medium, we measure a constant transmission, which is determined by $\Omega$ and the decoherence rate $\gamma_{gr}$. If we increase $R_{in}$, we observe a decay of transmission over time which gets faster with increasing rate $R_{in}$.  We attribute this effect to interaction-induced coupling to degenerate Zeeman sublevels which leads to polaritons being converted to stationary Rydberg excitations inside the cloud. These impurities shift the Rydberg levels of the surrounding atoms, preventing other polaritons from propagating through the cloud \cite{Duerr2014, Hofferberth2014, Rempe2014b}.

\begin{figure}[t!]
\begin{center}
\includegraphics[width=\columnwidth]{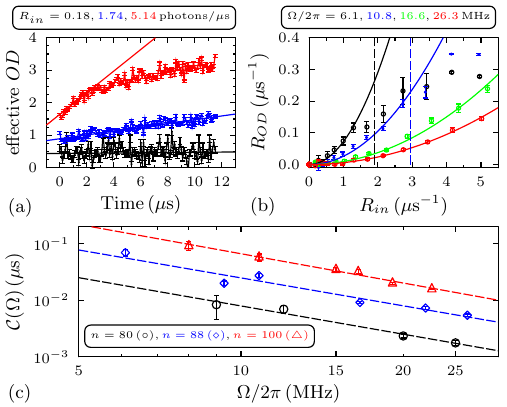}
\caption{\label{figure2} ~(a) Effective optical density of the  $|88D_{5/2}\rangle$ EIT feature for different photon rates at fixed $\Omega/2\pi = \unit[8.3] \mega \hertz$. The solid lines are linear fits to the data to extract the rates $R_{OD}$ of additional $OD_{e\!f\!f}$ due to the creation of impurities. (b) Dependence of $R_{OD}$ on $R_{in}$ for different Rabi frequencies $\Omega$. Parabolic fits (solid lines) determine the rate constant $\mathcal{C}(\Omega)$ for each dataset. The vertical dashed lines correspond to mean photon numbers inside the medium exceeding 2 where the quadratic dependence breaks down. (c) Rate constants $\mathcal{C}$ for different $\Omega$. For the different principal quantum numbers n we observe the same scaling according to Eq.\,\eqref{eq:OmegaDependence} with $k=1.67(4)$. Dashed lines are fits to the data.}
\end{center}
\end{figure}

To investigate the dependence of the decay of transmission on the photon rate $R_{in}$, we calculate an effective optical depth ($OD_{e\!f\!f}$) of our medium on EIT resonance $(\Delta = 0)$ by taking the logarithm of the transmission (Fig.\,\ref{figure2}(a)). To account for the three effects described above, we write  $OD_{e\!f\!f}$ as
\begin{equation}
OD_{e\!f\!f} = OD_{dec} + OD_{nl}(R_{in}) + OD_{dph}(R_{in},t),
\label{eq:TotalOptDensity}
\end{equation}
where the terms $OD_{dec}$, $OD_{nl}$ and $OD_{dph}$ represent the contributions of decoherence, blockade-induced nonlinearity, and dephasing. In the following we investigate the last contribution. Neglecting saturation effects for the highest values of $R_{in}$, we approximate the increase in $OD$ due to dephasing to be linear in time and write $OD_ {dph}=R_{OD}\cdot t$, with $R_{OD}$ as creation rate of optical density by decoupled impurities.

\begin{figure}[t]
\begin{center}
\includegraphics[width=\columnwidth]{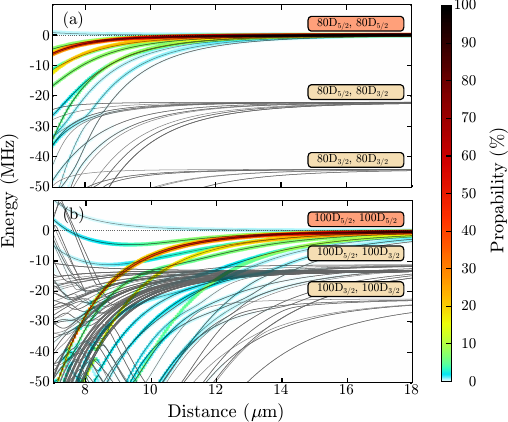}
\caption{\label{figure3} ~ Calculated pair state potentials of (a) $|80D_{5/2};80D_{5/2}\rangle$  and (b) $|100D_{5/2};100D_{5/2}\rangle$ pair states (gray lines). The color shading represents the projection of the $|m_{J1,2}=5/2,5/2\rangle$ state with respect to the the light propagation direction onto the new eigenstates in the presence of interaction for an angle of $\theta = 60^\circ$ between the interatomic axis and the coupling laser beam.}
\end{center}
\end{figure}

In Fig.\,\ref{figure2}(b) the extracted rates $R_{OD}$ are plotted versus $R_{in}$ for measurements with different control field Rabi frequencies $\Omega$ coupling to the $|88D_{5/2},m_J=5/2\rangle$ state. For measurements with $\Omega/2\pi = \unit[16.6] \mega\hertz$ and $\Omega/2\pi = \unit[26.3] \mega\hertz$, respectively, $R_\text{OD}$ scales quadratically with $R_{in}$ over the probed range. For  $\Omega/2\pi = \unit[6.1] \mega\hertz$ and  $\Omega/2\pi = \unit[10.8] \mega\hertz$ we see deviations from the quadratic dependence for $R_{in}$ exceeding $1.5$ and $2.7$, respectively. Considering the delay time in the medium given by $\tau_{delay} = \frac{OD\cdot\gamma_e}{\Omega^2}$ (where $\gamma_e/2\pi=\unit[6.1]\mega \hertz$ is the decay rate of intermediate state $|5P_{3/2}\rangle$) and finite initial EIT transmission at $t=0$, the onset of these deviations seems to coincide with rates corresponding to a mean number of photons in the medium exceeding 2 (vertical lines in Fig.\,\ref{figure2}(b)). This quadratic dependence of $R_{OD}$ on $R_{in}$ suggests that the observed dephasing is a two-body effect. In particular, we will later relate the dephasing to the probability $|\psiSS|^2$ of finding two polaritons in the Rydberg state at the same time for distances larger then the blockade radius $r_b$, given by
\begin{equation}
|\psiSS|^2 = \frac{R^2_{in}}{v^2_g}.
\label{eq:PolaritonAmplitude}
\end{equation}
Because of this dependence, we introduce a rate constant $\mathcal{C}(\Omega)$, relating $R_{OD}$ and $R_{in}$ in the quadratic regime via $ R_{OD} = \mathcal{C}(\Omega)\cdot R_{in}^2$, which we obtain from fits to our data (Fig.\,\ref{figure2}(b)).
%
%=================%
%  Figure 2       %
%=================%
%================================%
%
The $\Omega$ dependence of the extracted rate constants is shown in Fig.\,\ref{figure2}(c) for measurements with principal quantum numbers $n=80$, $n=88$ and $n=100$. To compare the results for different principal quantum numbers, we extract the dependence on $\Omega$ by a fit of the form
\begin{equation}
\mathcal{C}(\Omega) = a\cdot \Omega^{-k},
\label{eq:OmegaDependence}
\end{equation}
which yields $k=1.67(4)$ for all the different $n$. However, for the prefactor $a$ we observe a strong scaling with $n$, indicating significantly larger dephasing for larger $n$.
%
%=================%
%  Potentials     %
%=================%
In order to connect our observations to the anisotropy of the Rydberg interaction we calculate Rydberg pair potentials through diagonalization of the electrostatic DD interaction Hamiltonian \cite{Shaffer2006}. To reduce computation time we diagonalize the Hamiltonian in the coordinate system aligned with the atomic separation, where the total magnetic moment $M=m_{J1}+m_{J2}$ is conserved. Calculated pair potentials for $|80D_{5/2};80D_{5/2}\rangle$ and $|100D_{5/2};100D_{5/2}\rangle$ are shown in Fig.\,\ref{figure3}(a) and (b). The anisotropic DD interaction couples states with different magnetic quantum numbers in the $D$-pair state manifold, resulting in the observed splitting of the manifold in addition to the overall level shift. When considering the coupling of a laser field with given direction and polarization one has to account for the electric dipole selection rules by calculating the overlap of the coupled pair state in the fixed coordinate system defined by the direction of light propagation, with the eigenstates in the presence of interaction in the interatomic coordinate system \cite{Walker2008}. The colored shadow of the potentials shows the overlap of the $|nD_{5/2}, m_J = 5/2;nD_{5/2}, m_J =5/2\rangle$ pair state in the fixed coordinate system, rotated into the interatomic frame via Wigner d-matrices, with the new eigenstates for an angle $\theta=60^\circ$ between the interatomic axis and the light propagation direction. It can be seen that the coupled state is projected onto multiple new eigenstates. Unlike for isotropic $S$-states this projection depends strongly on the angle $\theta$ \cite{Walker2008}. One consequence of the anisotropic interaction is that the blockade volume for $D$-states is not spherical as the interaction potential can no longer be described by a single $C_6$ coefficient. Secondly, the coupled pair state is not an eigenstate in the presence of interaction and will thus experience a time evolution under the influence of the interaction Hamiltonian.

%=================%
% Figure 3        %
%=================%
\begin{figure}[htp]
\includegraphics{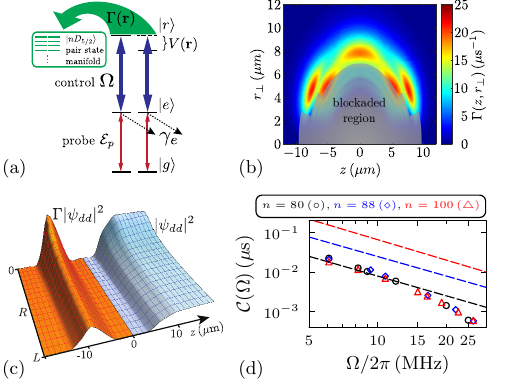}
\caption{(a) We model the anisotropic DD interaction between two polaritons as an effective potential $V(\br=\{\relative,r_{\perp}\})$ capturing the interaction-induced blockade effect, and a position dependent dephasing rate $\GamDph (\br)$ out of the pair-state coupled by the EIT control laser. (b) Example plot showing the anisotropic blockaded volume and the dephasing rate outside this region for $n=80$ and $\Omega/2\pi = 25\, \text{MHz} $. (c) Numerical simulations showing: For $z>0$ probability distribution associated with two Rydberg excitations $|\psiSS(z,R)|^2$. For $z<0$ the product $\GamDph(z)|\psiSS(z,R)|^2$, where $R=(z_1+z_2)/2$ is the center of mass coordinate. Both quantities are presented in arbitrary units for $r_{\perp}=4.2\mu m, \O/2\pi=12\, \text{MHz}, n=80$ and probe-beam waist $w_{\textit{eff}}=7\mu m$. (d) Comparison of $\mathcal{C}(\O)$ from numerical simulations (circles) with fits to experimental data from \figref{figure2}(c) (dashed lines).
}
\label{fig4}
\end{figure}

%================================%
%
%=================%
%  Theory         %
%=================%
To incorporate the interaction induced dynamics with polariton propagation under EIT, we reduce the full interaction potential to an effective level structure as shown in Fig.~\ref{fig4}(a). In the weak-probe limit, it is sufficient to consider only two photons simultaneously inside the medium \cite{Vuletic2012,Buechler2014}. Accounting for level shifts and selection rules, we determine the anisotropic blockade distance $r_b (\theta)$ where the laser-coupled pair state is shifted out of the EIT bandwidth (Fig. \ref{fig4}(b)). From this we obtain an effective interaction potential between two slow light polaritons $V(z,r_{\perp})$. This term results in the blockade-induced nonlinearity of the medium. Outside the blockaded region, where both the overall shift and the splitting of the $D_{5/2}$ Zeeman-manifold are smaller than the EIT bandwidth, we describe the evolution of the coupled pair-state by an effective dephasing rate $\GamDph(z,r_{\perp})$ of the Rydberg polaritons into localized Rydberg excitations. Here, $z=z_1-z_2$ and $r_{\perp}=r_{\perp,1}-r_{\perp,2}$ are the relative coordinate along and perpendicular to the propagation direction. To obtain $\GamDph(z,r_{\perp})$ we calculate the time evolution of a stationary Rydberg pair, in the initial state coupled by the fixed control field, at given distance and angle between light propagation direction and interatomic axis from the full interaction potential. Although this dynamics is fully coherent, the revival of the initial population appears only on time scales long compared to the polariton propagation time due to the large number of states in the $D$-state manifold. Hence, the initial time evolution on experimentally relevant time scales is well described by a spatially varying dephasing rate. The important result is that for $D$-states $\GamDph(z,r_{\perp})$ is large at distances well beyond the blockade volume (Fig. \ref{fig4}(b)). In contrast, the same approach results in vanishingly small dephasing rates for $S$-states.

For the propagation dynamics we solve numerically the full set of propagation equations for the two-body wave function \cite{Vuletic2012,Vuletic2013b}, including the dephasing rate $\GamDph(\relative,r_{\perp})$ as a decay of the amplitude $\psiSS(\br_1,\br_2)$ of two Rydberg excitations. We assume a homogeneous distribution of atoms inside the finite-size medium of length $L=4\s_z$ and include imperfect single-photon transmission due to the decoherence $\gamma_{gr}$ of the 2-photon transition. Neglecting probe-beam diffraction due to the interaction, we solve polariton dynamics (\figref{fig4}(c)) for different $r_{\perp}$ with $\vOneD(z)=V(z,r_{\perp})$ followed by averaging over the $r_{\perp}$ distribution determined by the Gaussian transverse profile with waist $w_{\textit{eff}} = 7\mu m$ (corresponding to the waist averaged over length $L$) of the probe beam.
%
%=================%
%  Figure 4       %
%=================%
%================================%
%
Then, the rate of events $\nD$ that at least one photon is converted into stationary Rydberg excitation is proportional to the amplitude of the two-polariton wavefunction $\psiSS$:
\begin{eqnarray}
 \nD &=& \int_{V} d{\bf r_1} d{\bf r_2}   \GamDph({\bf r_1}-{\bf r_2}) |\psiSS({\bf r_1}, {\bf r_2})|^2 \rm.
\label{avgRates}
\end{eqnarray}
We normalize the Rydberg wave function $\psiSS$ with the incoming photon flux as in Eq.\,(2). A single dephasing event increases on average the optical depth within the medium  by  $OD_{im}$ for the incoming photons. Therefore, the change in transmission by a single dephasing event  is  $\exp({- OD_{dec}-OD_{sat}(R_{in})})\left(1-e^{-OD_{im}}\right)$. This reduction appears with the rate $\nD$,
and therefore the initial time evolution for the averaged transmission behaves as
\begin{equation}
 T(t) = e^{- OD_{dec}-OD_{sat}(R_{in}) } \exp\left[- \nD t \left(1 - e^{- OD_{im}}\right)\right]\nonumber
\end{equation}
and  leads to a rate constant $\mathcal{C}(\O)=\nD\left(1 - e^{- OD_{im}}\right)/R_{in}^2$.
Furthermore, we observe, that a finite life time of the Rydberg impurity results in an effective saturation of the transmission. However, the full time evolution for the transmission including higher number of excited Rydberg impurities is extremely challenging and beyond the scope of the present manuscript. These calculations provide important insights into the behavior of the dephasing. First, the averaged optical thickness $OD_{im}>1$ of a
dephasing event is sufficient to strongly block the medium. Therefore, the decay mainly follows the probability to absorb at least one impurity. Second, there appears a characteristic distance with the highest probability for the excitation of an impurity Rydberg state given by the competition between the higher dephasing rates at shorter distances and the suppression to find two Rydberg excitations due to the blockade effect, \figref{fig4}(b-c).
The latter is strongly affected by the polariton dynamics inside the medium, as has been previously discussed in terms of a diffusive behavior \cite{Vuletic2012}: at the entrance of the two photons into the
medium, the probability to find two Rydberg excitations is purely determined by the blockade due to interactions $\psiSS(z) \sim 1/(1-\bar{\chi} \vOneD(\relative))$ where $ \hbar \bar{\chi} = -i({\GamP}/{ \Omega^2} +{1}/{ \GamP})$ \cite{Buechler2014}. This dip in probability broadens during propagation due to correction to the linear behavior of the polariton dispersion, while the single polariton losses provide an overall reduction of the amplitude (see \figref{fig4}(c)). These effects strongly depend on $\O$. In addition, photons inside the medium are compressed due to the slow light velocity, which contributes an additional factor $\O^{-4}$ (Eq.\,\eqref{eq:PolaritonAmplitude}) to the scaling of  $\mathcal{C}(\O)$ with $\O$. Both described effects combined explain the numerical results presented in \figref{fig4}(d).
We find qualitative agreement between theory and experiment without any fitting parameters. While for low $n$ and small values of $\O$ the agreement is excellent, for larger $\O$ we observe a discrepancy, and moreover, the theory does not reproduce the strong scaling with main principal quantum number $n$ measured in the experiment. We expect the reason for this discrepancy to be the fact that for large $\O$ and $n$ the AC-stark shift and broadening of the Rydberg lines, caused by coupling to the $5P_{3/2}$ manifold by the control field, becomes comparable to the splitting between the $nD_{5/2}$ and $nD_{3/2}$ manifold which scales with $n^{-3}$. In this case, our two-step approach of first calculating the interaction potentials and then incorporating them into the polariton propagation does not capture the full evolution of the system.

%================%
%Conclusion and Outlook%
%================%
In summary, we have investigated the effect of anisotropic DD interaction between Rydberg $D$-state polaritons. Interaction-induced decoupling into stationary Rydberg excitations results in a decrease of transmission on EIT resonance over time. This effect is relevant to all Rydberg experiments employing non-$S$-states \cite{Adams2013} or F\"{o}rster resonances \cite{Weidemueller2013b, Rempe2014b}, where the anisotropy of the Rydberg interaction will always result in coupling to additional levels. Our theoretical approach to include full Rydberg pair state potentials in numerical two-photon propagation in the form of an effective potential and anisotropic decay rate yields qualitative agreement with our measurements and thus is a useful tool for treating the complicated interaction between Rydberg polariton and stationary excitations in general. The fact that we observe the interaction-induced state-mixing on the few-photon level is a promising result for experiments on Rydberg-dressing \cite{Zoller2015b,Pohl2015} and engineered polariton-interaction \cite{Gorshkov2015} using Rydberg states with orbital angular momentum. More detailed study of the anisotropic coupling will be possible in storage and retrieval experiments \cite{Duerr2014, Kuzmich2012c}, enabling control over number and position of stored excitations. In this scenario, it becomes particularly interesting to employ echo techniques \cite{Kimmich2001} to probe the coherent spin evolution of interacting Rydberg-polaritons.

We thank Rick van Bijnen for important input in developing the core idea of this work and Thomas Pohl for helpful discussions. This work is funded by the German Research Foundation (DFG) through Emmy-Noether-grant HO 4787/1-1 and within SFB/TRR21. HG acknowledges support from the Carl-Zeiss Foundation, PB from EU Marie Curie ITN COHERENCE.

%\bibliographystyle{apsrev4-1}
%\bibliography{biblio_dstate_paper}

%\begin{thebibliography}


%\end{thebibliography}

\end{document}